\documentstyle[epsfig,12pt]{article}

\def\Journal#1#2#3#4{{#1} {\bf #2}, #3 (#4)}

\def\NPA{{Nucl. Phys.} A}
\def\NPB{{Nucl. Phys.} B}
\def\PLB{{Phys. Lett.}  B}

\def\PRL{Phys. Rev. Lett.}
\def\PRC{{Phys. Rev.} C}
\def\PRD{{Phys. Rev.} D}

\newfam\BMath
\font\BMathL=cmmib10 
\font\BMathl=cmmib7
\font\BMathm=cmmib5
\textfont\BMath=\BMathL \scriptfont\BMath=\BMathl
\scriptscriptfont\BMath=\BMathm

\def\a{\alpha}

\def\e{\epsilon}

\def\q{\theta}

\def\t{\tau}

\def\be{\begin{equation}}
\def\ee{\end{equation}}
\def\bea{\begin{eqnarray}}
\def\eea{\end{eqnarray}}
\def\eref#1{Eq.~(\ref{#1})}

\def\fref#1{Fig.~\ref{#1}}
\def\bfi{\begin{figure}}
\def\efi{\end{figure}}

\newcommand{\ncom}{\newcommand}
\ncom{\vo}[1]{{\fam\BMath #1}}
\ncom{\vt}[2]{({\fam\BMath #1}+{\fam\BMath #2})}
\ncom{\vmo}[1]{\vert{\fam\BMath #1}\vert}
\ncom{\vmt}[2]{\vert\mbox{\bf #1}+\mbox{\bf #2}\vert}
\ncom{\lan}{\langle}
\ncom{\ran}{\rangle}
\ncom\nonum{\nonumber \\}
\ncom\fx{\!\!\!\!}
\ncom\gsim{\mbox{\raisebox{-0.6ex}{\ $\stackrel {>}{\sim}$\ }}}
\ncom\lsim{\mbox{\raisebox{-0.6ex}{\ $\stackrel {<}{\sim}$\ }}}

\ncom{\half}{{1\over 2}}
\ncom{\third}{{1\over 3}}
\ncom{\fourth}{{1\over 4}}
\ncom{\fifth}{{1\over 5}}
\ncom{\sixth}{{1\over 6}}

\ncom\Tg{T_{eq\; g}}
\ncom\Tq{T_{eq\; q}}
\ncom\qg{\q_g}
\ncom\qq{\q_q}

\value{textfraction}

\begin{document}

\begin{flushright}
\footnotesize \sffamily WU B 97/16
\end{flushright}

\vskip 3.0truecm

\begin{centering}
{NON-ABELIAN EQUILIBRATION IN HEAVY ION COLLISIONS}

\null
\null

{S.M.H. Wong}

\em Fachbereich Physik, Universit\"at Wuppertal,
D-42097 Wuppertal, Germany

\end{centering}

\vskip 6.0cm

\begin{abstract}

The uncertainty in the equilibration in heavy ion 
collisions due to the choice of the value of the 
coupling is examined. The results of the equilibration
are indeed affected by this choice. In particular, a 
variation of $\a_s$ from 0.3 to 0.5 reduces the parton phase 
of the plasma by as much as 4.0 fm/c and increasing 
coupling causes a reduction in the generated entropy. 
As far as equilibration is concerned, larger coupling
results in faster equilibration both chemically
and kinetically but improvements only happen to
the fermions. These are accompanied by more
rapid cooling and therefore shortened lifetime.
In the light of these results, to choose $\a_s$ is
almost equivalent to choosing the results. Also because of
consistency, any fixed $\a_s$ is incompatible 
with an evolving plasma. The best choice is therefore 
not to choose at all but let the system makes its
own decision. We show a simple recipe how this can
be done. With an evolving coupling, equilibration is
accelerated with better results at the expense of
the duration of the deconfined phase.

\end{abstract}

\vfill
\break

\baselineskip 17pt

\section{Introduction}
\label{sec:intro}

In future heavy ion collision experiments at LHC
and at RHIC, a major effort is not only to try to
produce deconfined matter, given some of which
might have already been created at CERN SPS 
at present energies, but also to show beyond any doubt
that during the collisions that deconfined matter
really exists for however brief moment. At AGS 
and SPS, this latter task would be difficult because
their energies might not be sufficiently high or the 
system might not be large enough to reduce boundary 
effects on the produced quark-gluon plasma. 

In the very violent collisions, many particles 
will be produced, some of which can escape from
the system such as electromagnetic probes,
others can only reveal themselves after the final
break up of the system such as probes using
strange and charm hadrons. Nevertheless, they
will all be affected by the evolution of the
system. Particle production are not exactly the
same during the initial, pre-equilibrium and 
equilibrium phases, therefore the durations of these
periods also play a role in modifying the yield of the 
final produced particles which will eventually fall into
the detectors. It is therefore necessary to study
the evolution of the system or equivalently to 
study the equilibration of the parton plasma. 
There have already been quite a few works on
the equilibration in the parton phase of the
plasma, for example chemical equilibration 
$^{1)}$, thermalization $^{2)}$ and full 
equilibration $^{3) 4)}$. 
Our purpose here is not exactly to repeat these
investigations but to concern ourselves on one of
the uncertainties which can alter the results of
the equilibration. The first and most obvious 
is caused by the uncertainty in the initial inputs. 
These can be studied but we will not do this at present.
The second uncertainty could have been the infrared 
screening parameter used in the usual perturbative QCD.
However, in a multi-particle system, QCD will generate
Debye, quark and gluon medium masses $^{5)}$.
These will effectively screen off the infrared 
divergences. So this parameter and therefore the
uncertainty needs not be present. The third one is
the strong coupling constant itself. It is the only
remaining free parameter once that of infrared screening
is no longer present. We would like to find out how the 
results will change with the value of $\a_s$. There is 
however a second reason concerning consistency for which
we would like to look into the uncertainty arising
from the choice of $\a_s$. The common choice
for these kinds of studies is $\a_s=$ 0.3 which
corresponds to an average momentum transfer of
$Q \sim$ 2.0 GeV with $\Lambda_{QCD}=$ 200 MeV. 
This choice is reasonable early on after the initial
collisions. We show in \eref{fig:ave}, the evolution
of the average energy for quarks and gluons in our
previous investigation into the equilibration at
LHC and at RHIC. As can be seen, the average parton
energy drops by at least 1.0 GeV over the
entire evolution. Since the average momentum
transfer has to be related to the average parton 
energy, and so $\a_s=$ 0.3 cannot be a good value
for the entire duration of the evolution. In fact, any 
choice of fixed $\a_s$ will equally not be good
enough. We will show a recipe for obtaining a varying 
coupling in the next section to overcome this problem of
consistency and will show the corresponding results
as well as those with other fixed values of the $\a_s$.
Together they reveal the effects of the coupling on the
equilibration.

\bfi
\centerline{
\hbox{
\epsfig{file=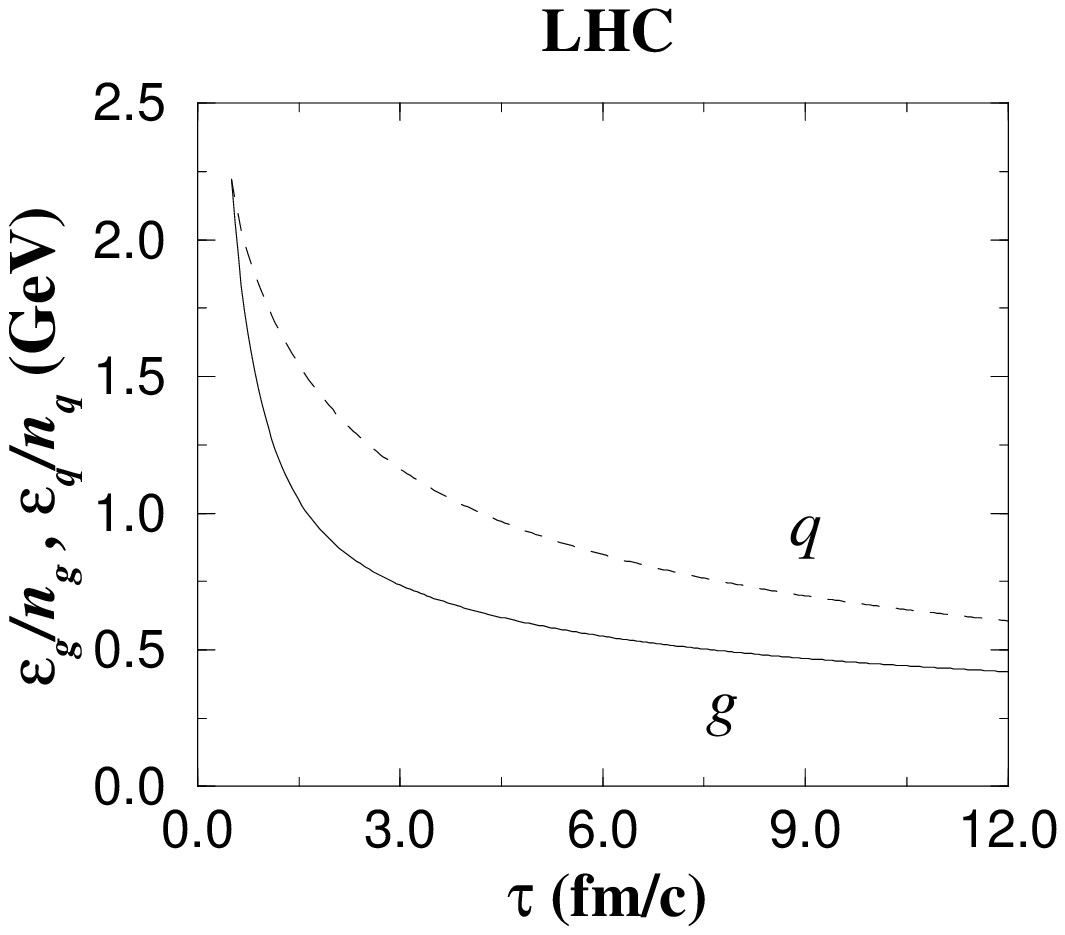,width=2.3in}\ \ \ \ \
\epsfig{file=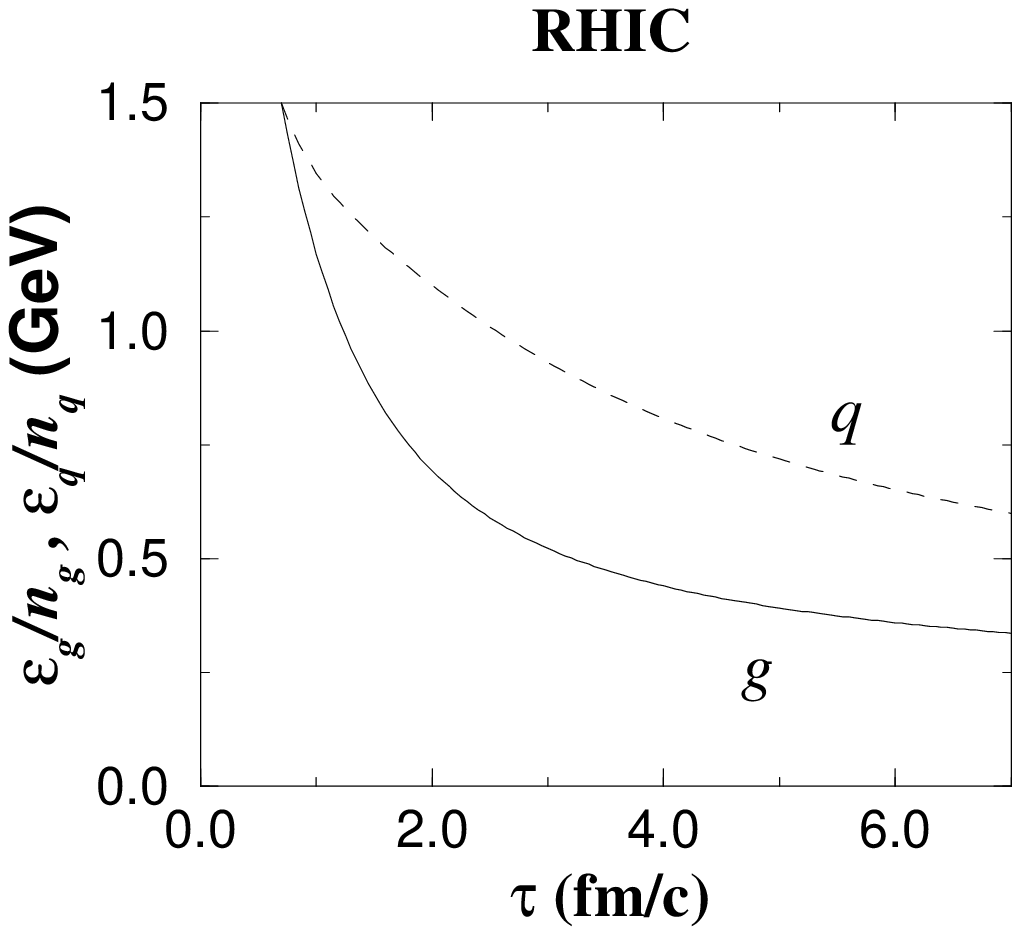,width=2.3in}
}}
\caption{\baselineskip 15pt
The average gluon (solid) and quark (dashed) energies 
change a lot during the evolution of the parton plasma 
both a LHC and at RHIC. Therefore the average momentum transfer
should also vary considerably and hence the coupling as well.}
\label{fig:ave}
\efi

\section{Evolution of a parton plasma with various
couplings}
\label{sec:evol}

To study and to make the effect of $\a_s$ manifest,
we choose some large values of $\a_s=$ 0.5, 0.8 to do
the evolution in addition to the previous
$\a_s=$ 0.3 case. Now also to overcome the problem of 
consistency, we introduce, as a solution, a varying 
coupling which evolves with the system in the following 
way. Since two incoming partons each carrying the 
average parton energy 
$\lan E \ran =\lan \e_{tot}\ran /\lan n_{tot}\ran$
can exchange a maximum $Q^2 =4 \lan E\ran^2$. So combining
the $\lan E(\t)\ran$ at any moment $\t$ and the 1-loop
running coupling formula, we have a coupling that is 
entirely determined by the system. With this latter
approach, there is no remaining free parameter.
This is in fact, with hindsight, a better choice for
the coupling as we will see presently. We denote
this coupling by $\a_s^v$ from now on. 

With these choices of the coupling, we show the 
effects of the coupling in \fref{fig:fugtem} and
\fref{fig:pres}. \fref{fig:fugtem} shows the
effects on chemical equilibration in terms of the
fugacities as well as the parton estimated 
temperatures. \fref{fig:pres} shows the pressure 
and energy density to pressure ratios from which 
we can deduce information on kinetic 
equilibration. Let us first look at parton chemical
equilibration. We used the same fixed
initial conditions as before $^{4)}$ so as to 
concentrate only on the coupling. From \fref{fig:fugtem},
we see that with increasing $\a_s$, chemical equilibration
is definitely faster both for gluons and for quarks,
however, only in the case of the fermion, do they
show any improvements. For gluons, the end degree of
chemical equilibration does not change very much with
the coupling.

\bfi
\centerline{
\hbox{
\epsfig{file=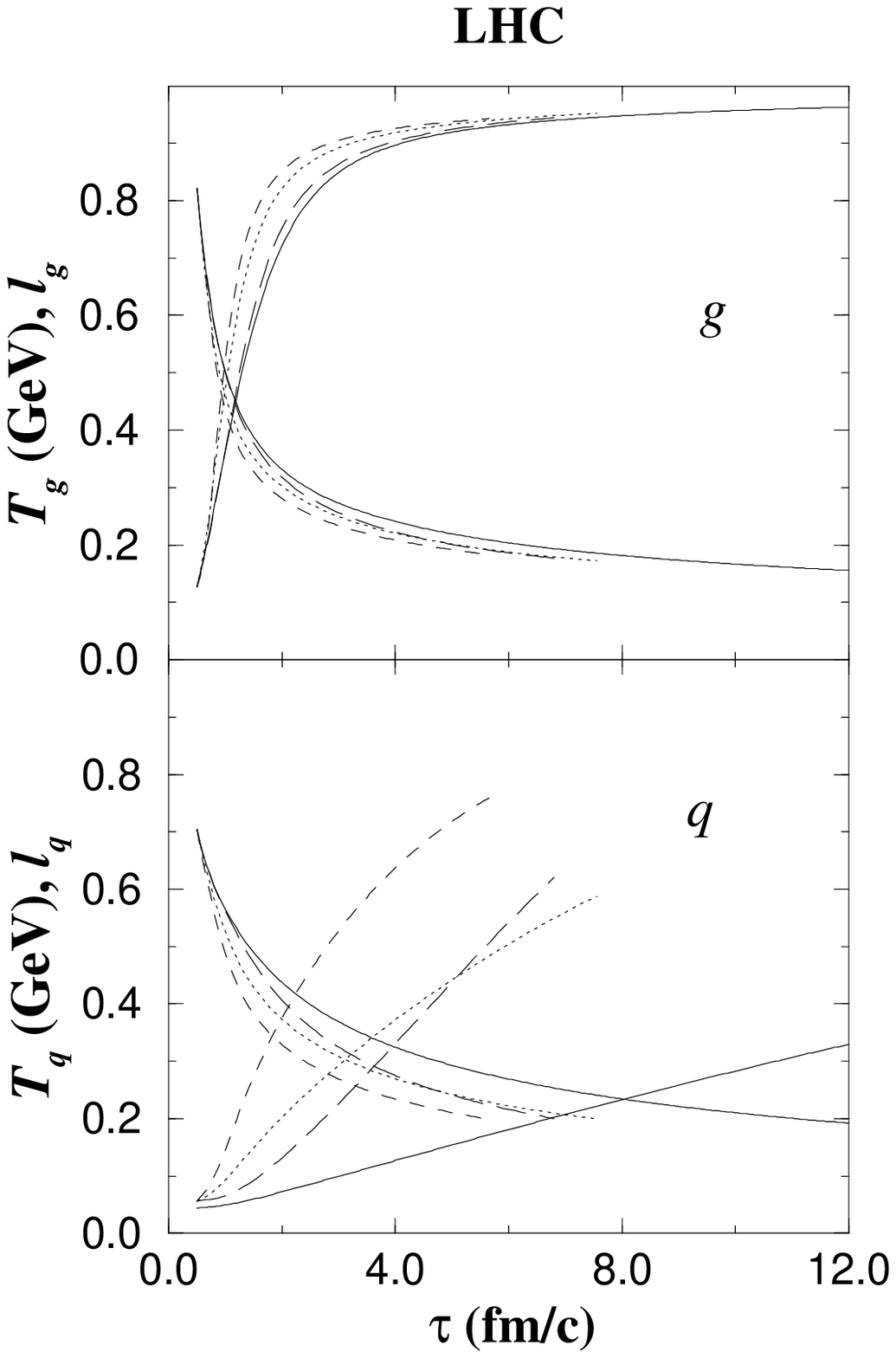,width=2.3in}\ \ \ \ \
\epsfig{file=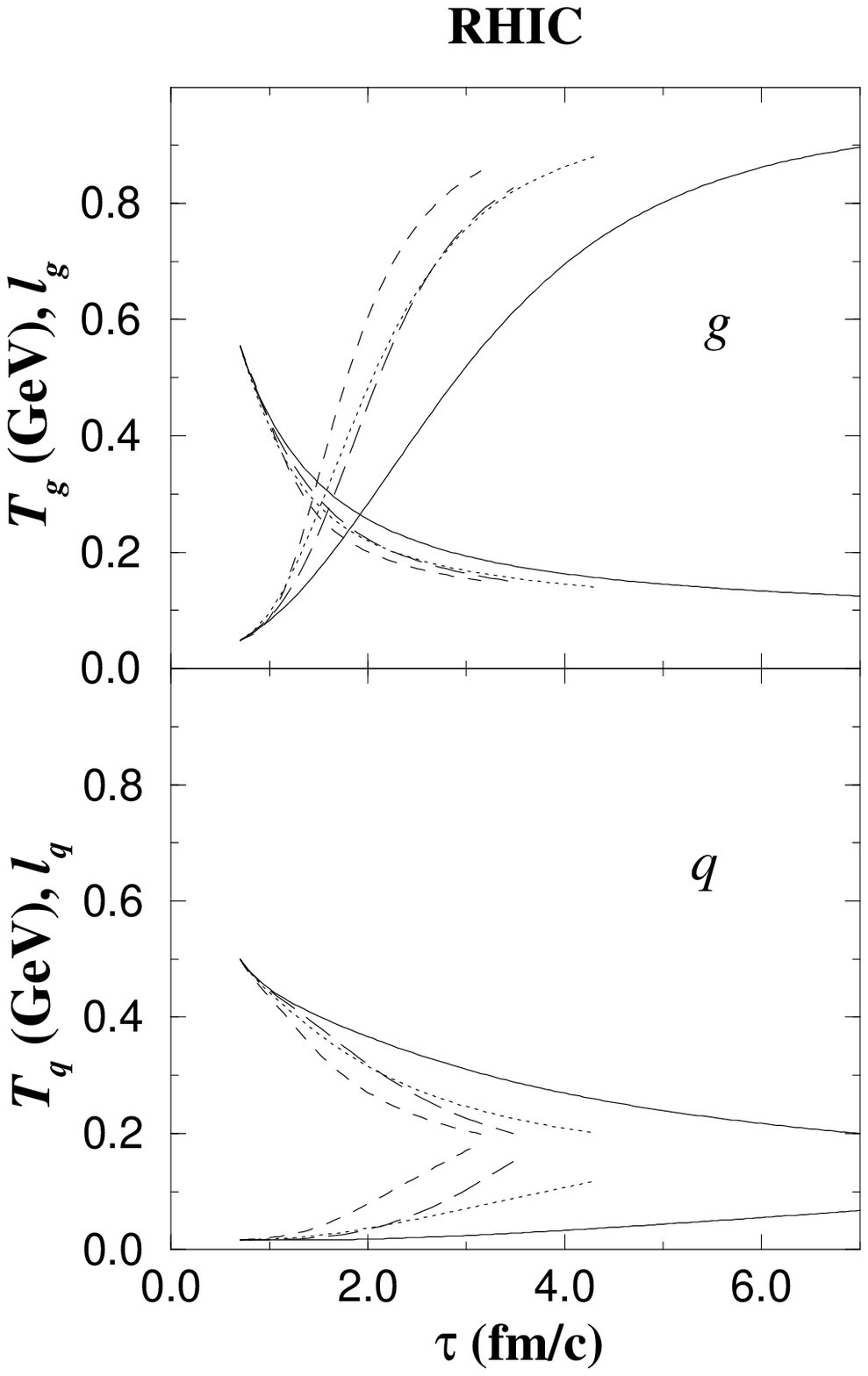,width=2.3in}
}}
\caption{\baselineskip 15pt
With increasing coupling, chemical equilibration
is faster but the cooling is also more rapid. As seen here, 
the fastest increase in the parton fugacities $l_g$ and $l_q$ 
are the curves (dashed) with $\a_s=0.8$, the next are those
produced with $\a_s=0.5$ (dotted) and $0.3$ (solid).
The curves of $\a_s^v$ (long dashed) shift across the constant
$\a_s$ ``contours'' and equilibrate definitely better than
the solid lines. The shifts of the descending $T$ curves
with $\a_s$ are sizable. As a result, the lifetime of the
parton phase of the plasma is controlled by the value of the
coupling.}
\label{fig:fugtem}
\efi

For kinetic equilibration, we use the pressure to
pressure and energy to pressure ratios as a check of the
isotropy of the parton momentum distribution.
Because of our thermalized initial conditions, these
ratios should start at 1.0 or with an isotropic distribution.
This isotropy is lost subsequently, as seen in \fref{fig:pres},
where all the curves shift downward away from 1.0. 
This is due to the disruptive effect of the longitudinal
expansion. As the net interaction responds by increasing
its rate, the expansion effect is later overcome and
isotropy is progressively being recovered. This is when
all the curves in \fref{fig:pres} rise again. Larger
couplings lead to faster equilibration but as in
the case of chemical equilibration, only quarks and 
antiquarks show obvious improvements. Equilibration
is clearly faster for all partons in \fref{fig:fugtem}
and \fref{fig:pres}. However, this is achieved at the
expense of similarly faster cooling. The temperature
estimates all decrease more rapidly than before in
\fref{fig:fugtem}. Not only that but the time at
which they reach the assumed phase transition 
temperature at $T_c \sim$ 200 MeV changes by as much as 
4.0 fm/c when the $\a_s$ varies from 0.3 to 0.5 at LHC.
So the lifetime of the parton phase of the plasma
is very sensitive to the value of $\a_s$. Similarly
sensitive to the coupling is the generated entropy
$^{4) 6)}$. It is clear therefore, as we have
already mentioned, to have to choose a value for $\a_s$
is not the best choice. With the consistent $\a_s^v$,
we see in \fref{fig:fugtem} and \fref{fig:pres},
these curves start off near $\a_s=$ 0.3 case but
shift away across the constant $\a_s$ ``contours''.
They therefore achieve faster and better equilibration
for quark and antiquark than the $\a_s=0.3$ case. They
equilibrate, in this case, in an accelerated fashion. 
This is a non-abelian effect not found in the
equilibration of ordinary electromagnetic plasma
or in other many-body system.

\bfi
\centerline{
\hbox{
\epsfig{file=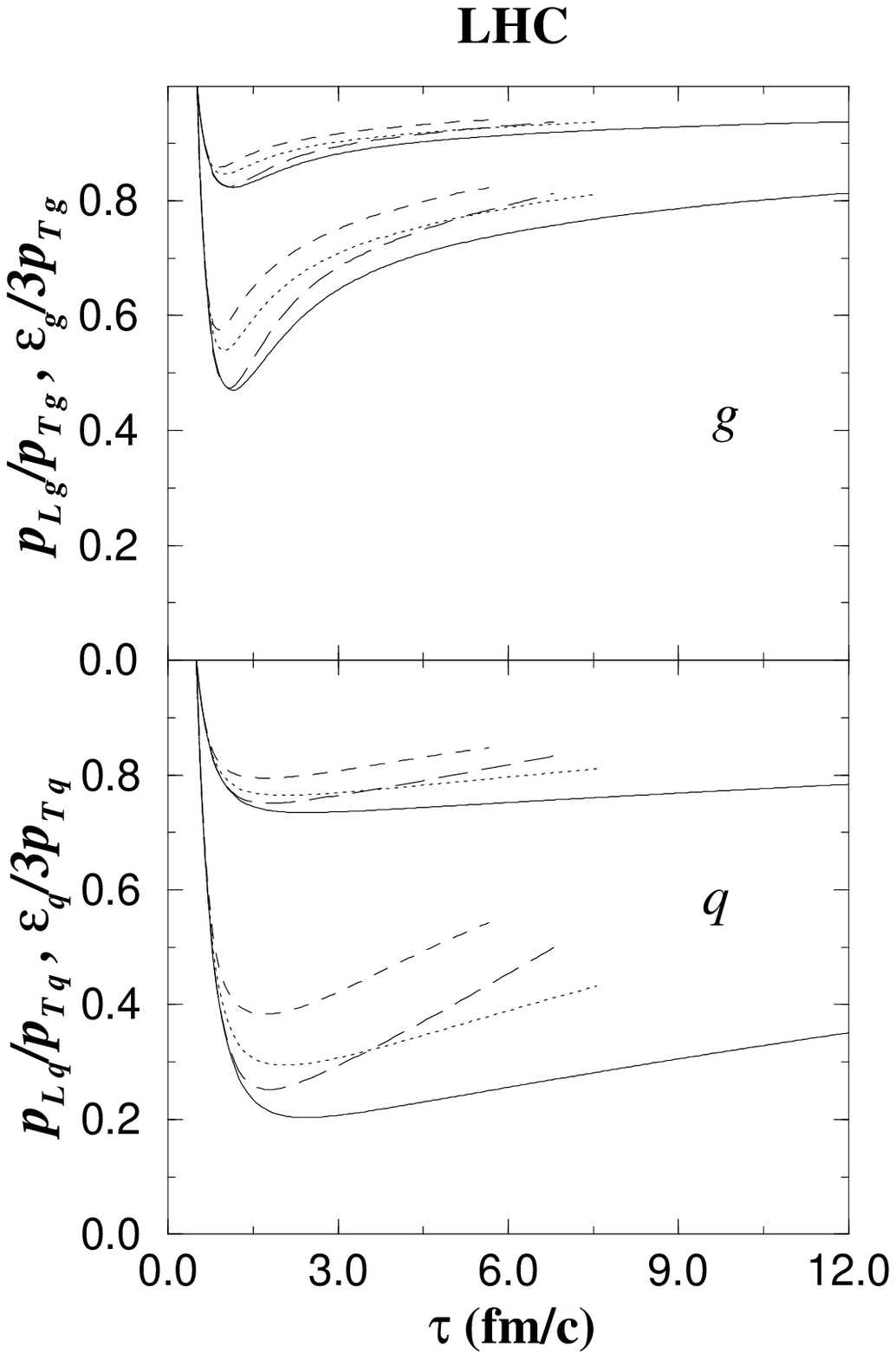,width=2.3in}\ \ \ \ \
\epsfig{file=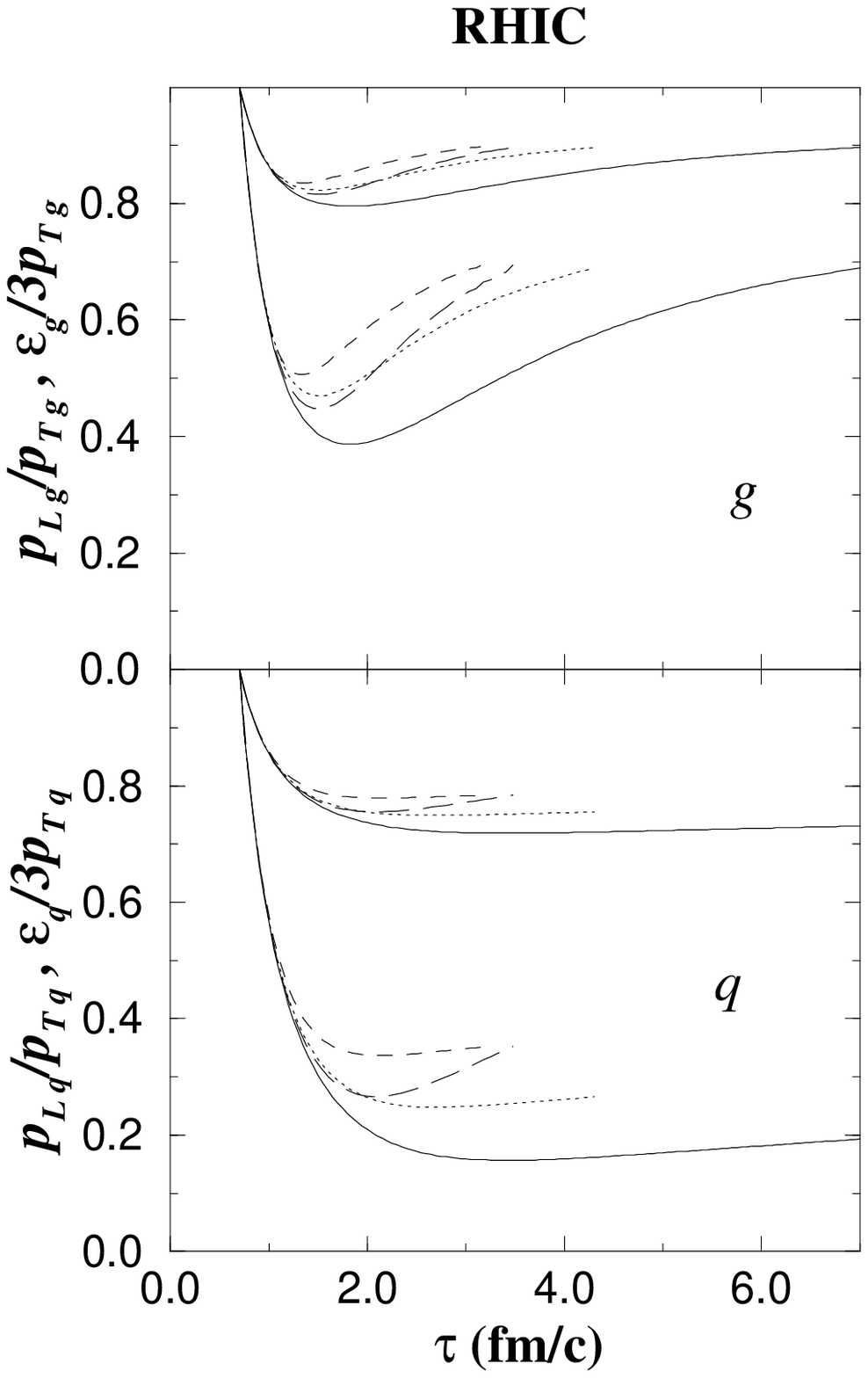,width=2.3in}
}}
\caption{\baselineskip 15pt
These ratios show the isotropy of the parton
momentum distribution and therefore kinetic equilibration
with $\t$. The top (bottom) set of four curves are for 
energy (pressure) to pressure ratios. Faster thermalization
is seen with increasing coupling. The assignment of the
coupling to the curves are $\a_s=$ 0.3 (solid), 0.5 (dotted),
0.8 (dashed) and $\a_s^v$ (long dashed). Improvements are,
however, reserved only for the fermions.}
\label{fig:pres}

\null
\centerline{
\hbox{
\epsfig{file=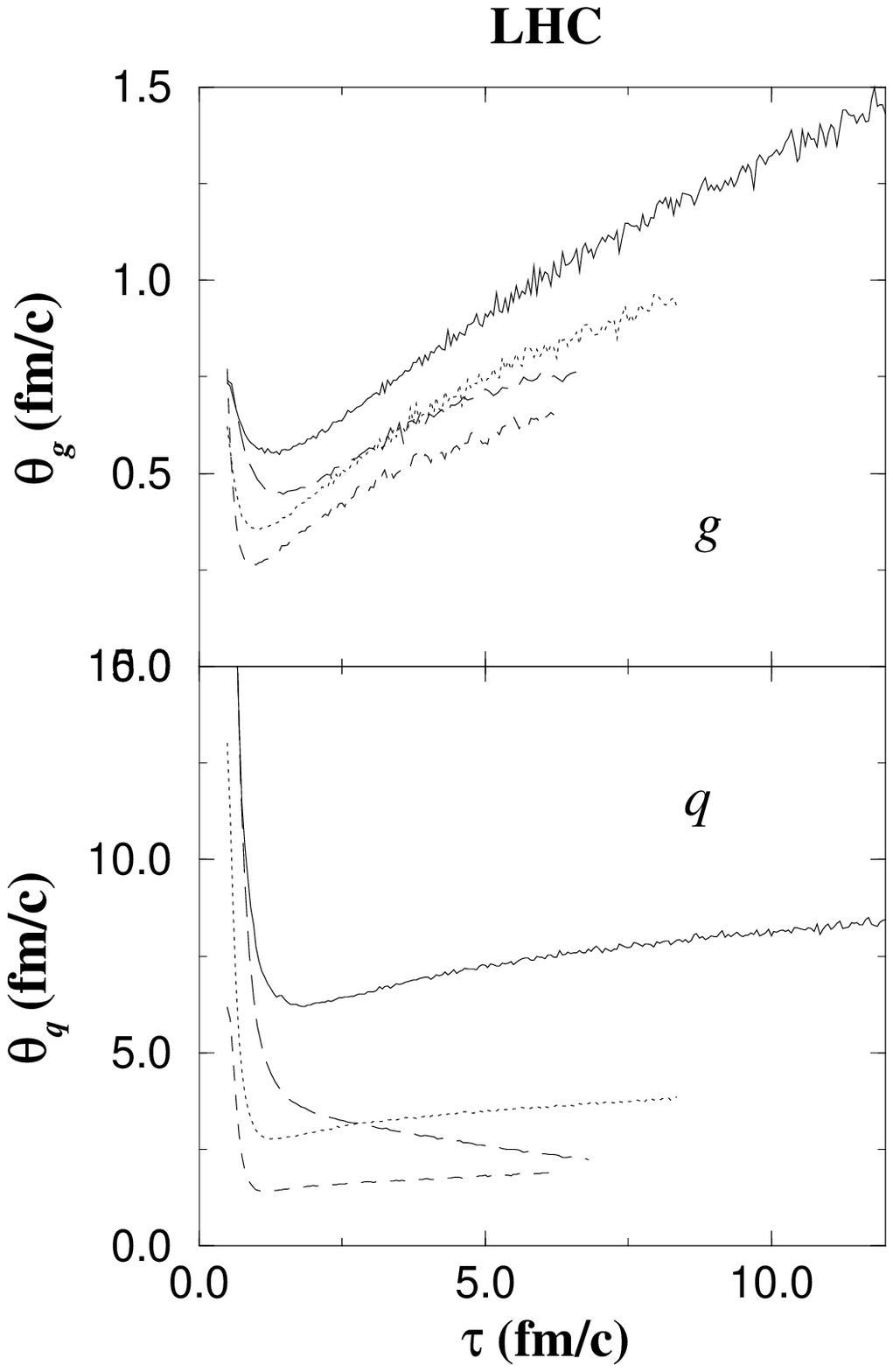,width=2.3in}\ \ \ \ \
\epsfig{file=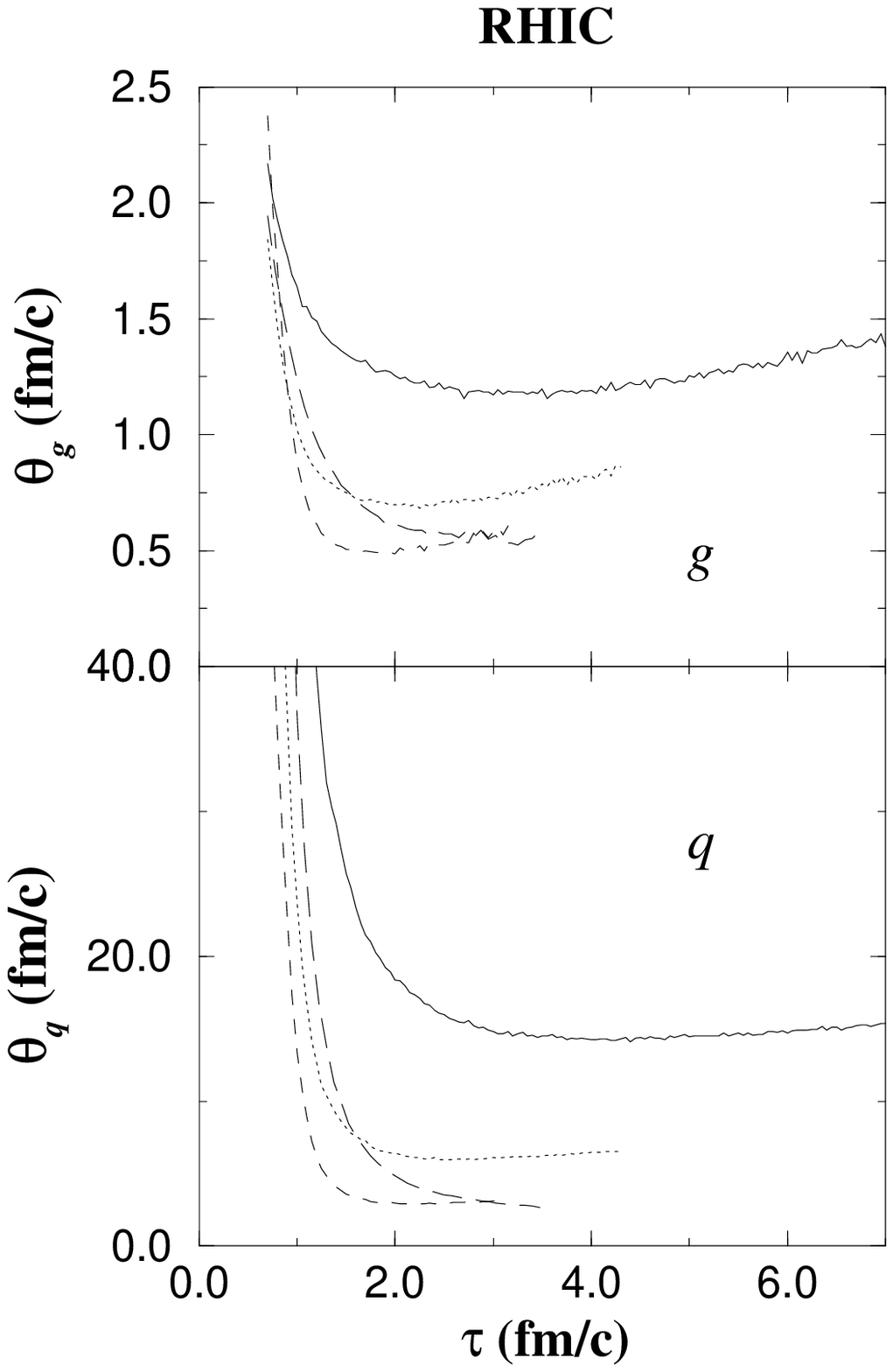,width=2.3in}
}}
\caption{\baselineskip 15pt
Evolution of the collision time reflects 
that of the net interaction rate. The values of the
coupling are assigned to the curves in the same way
as before. The curves produced with fixed couplings
have similar time-dependent behaviours. The case
of $\a_s^v$ is, however, very distinct. It shows the
non-abelian effect of QCD accelerates the 
equilibration. This is unique to a QCD parton plasma.
}
\label{fig:coll}
\efi

The non-abelian effect mentioned in the previous
paragraph can be shown more clearly via the collision
time $\q$ defined previously through the relaxation time
approximation $^{4) 6)}$. In \fref{fig:coll},
we have plotted the evolution of these for the various
values of the coupling. From the fixed coupling $\q$'s,
their behaviours are similar, i.e. a fast rapid initial
decrease and then a slow rise until the end. These
can be explained by the short expansion and the net
interaction dominated phases. It would be helpful to
identify the inverse collision time as the net interaction
rate, then the first phase is simply the response of the
system to being driven out of equilibrium. As a consequence,
the net interaction rate has to increase or $\q$ has to 
drop. As equilibrium is approached, the net rate has to
slow down so $\q$ has again to increase but slowly. 
In the case of $\a_s^v$, the slow rise of the net 
interaction dominated phase is replaced by a continued
slow decrease. So the net rate continues to increase
albeit at a much slower rate as the initial drop.
We have already seen in \fref{fig:fugtem} and
\fref{fig:pres} that equilibration is present in all 
cases, so what is happening to the $\a_s^v$ case is, the
increase in the interaction strength compensates for
the near equilibrium slowing down of the net rate. 
The result is the non-abelian accelerated equilibration
that we have already seen. This is of course unique
to a QCD plasma.

In summary, we have studied the $\a_s$ dependence of 
the equilibration and introduced a simple recipe to 
solve the consistency problem raised in the introduction.
It is found that larger $\a_s$ means faster
equilibration for all partons but improvements are only
for the fermions. Lifetime of the parton phase as well
as the entropy are sensitive to the value of $\a_s$
therefore the consistent $\a_s^v$ is a better choice
which gives rise to accelerated equilibration unique
to a QCD plasma.

\section*{Acknowledgements}

The author would like to thank the organizers 
for this interesting and informative conference.

\section*{References}

\begin{itemize}

\item[1)]T.S. Bir\'o, E. van Doorn, B. M\"uller, 
M.H. Thoma and X.N. Wang, \Journal{\PRC}{48}{1275}{1993};
P. L\'evai, B. M\"uller and X.N. Wang, \Journal{\PRC}{51}{3326}{1995};
X.N. Wang, \Journal{\NPA}{590}{47}{1995}; 
E.V. Shuryak and L. Xiong, \Journal{\PRC}{49}{2203}{1994}.

\item[2)]G. Baym, \Journal{\PLB}{138}{18}{1984};
S. Gavin, \Journal{\NPB}{351}{561}{1991};
K. Kajantie and T. Matsui, \Journal{\PLB}{164}{373}{1985};
E.V. Shuryak, \Journal{\PRL}{68}{3270}{1992};
H. Heiselberg and X.N. Wang, \Journal{\PRC}{53}{1892}{1996};
H. Heiselberg and X.N. Wang, \Journal{\NPB}{462}{389}{1996}.

\item[3)]K. Geiger and B. M\"uller, 
\Journal{\NPB}{369}{600}{1991}; K. Geiger, 
\Journal{\PRD}{46}{4965, 4986}{1992};
K. Geiger and J.I. Kapusta, \Journal{\PRD}{47}{4905}{1993};

\item[4)]S.M.H. Wong, \Journal{\NPA}{607}{442}{1996};
\Journal{\PRC}{54}{2588}{1996}.

\item[5)]V.V. Klimov, \Journal{Yad. Fiz.}{33}{1734}{1981},
\Journal{Sov. J. Nucl. Phys.}{33}{934}{1981}, A.H. Weldon, 
\Journal{\PRD}{26}{1394,2789}{1982}.

\item[6)]S.M.H. Wong, $\a_s$ dependence in the 
equilibration in relativistic heavy ion collisions, 
preprint WU B 97/13.

\end{itemize}

\end{document}